\documentstyle[aps,prl,epsfig,multicol]{revtex}
\begin{document}
\draft
\title{Effect of fluctuations on vortex lattice structural transitions in superconductors}
\author{A. Gurevich} 
\address{Applied Superconductivity Center 
University of Wisconsin, Madison, Wisconsin 53706} 
\author{V.G. Kogan}
\address{Ames National  Laboratory and Department of Physics and Astronomy, Iowa State University, 
Ames Iowa 50011}
\date{\today}
\maketitle
\begin{abstract}
 The rhombic-to-square transition field $H_{\Box}(T)$  for cubic and
tetragonal materials in fields along [001] is evaluated using the nonlocal
London theory with account of thermal vortex fluctuations. Unlike extended
Ginzburg-Landau  models, our approach shows that the line $H_{\Box}(T)$
and the upper critical field $H_{c2}(T)$ do not cross
 due to strong fluctuations near $H_{c2}(T)$ which suppress the
square anisotropy induced by the nonlocality. In increasing fields, this
causes  re-entrance of the rhombic structure   in agreement with recent
neutron scattering data on borocarbides.   
\end{abstract}
\begin{multicols}{2}

Complex behavior of vortex lattices (VLs) even in cubic superconductors
like Nb has been known for a long time \cite{Nb}. Recent  progress in
understanding  the evolution of VLs with the magnetic  field $H$ and
temperature $T$ became possible due to availability of large high quality
crystals of borocarbides \cite{Paul}, which  triggered a score of
small-angle neutron scattering (SANS) \cite{n4,n7,n8}, scanning tunneling
\cite{n5,sakata}, and decoration experiments \cite{vinn}.  
Of a particular interest  is the ubiquitous structural transition
between rhombic and square VLs in increasing   fields   along the
$c$ axis 
of tetragonal borocarbides \cite{n4,n7,n8}. 

For this case, the vortex repulsion is {\it isotropic} within the
standard London and  Ginzburg-Landau (GL) theories, so that  vortices
should always form the hexagonal Abrikosov lattice, which provides the
maximum vortex spacing for a given flux density $B/\phi_0$ ($\phi_0$ is
the flux quantum).  There is no coupling between the VL and the crystal 
in  these models; as a consequence, the VL orientation is arbitrary and no
VL structural transitions  are expected.   A full microscopic
theory of the mixed state contains this coupling, but involves   
self-consistent calculations of the  gap and current distributions, a
formidable task even for materials with the GL parameter
$\kappa\sim 1$ \cite{micro}.  The situation simplifies in  high-$\kappa$ 
materials (like borocarbides), for which one can utilize a more
transparent nonlocal London  model\cite{nl}. Within this approach, the VL 
coupling to the crystal is provided by the basic nonlocal
relation between the current density and the vector potential,
$J_{\alpha}({\bf r})=
\int Q_{\alpha\beta}({\bf r}-{\bf r}')A_{\beta}({\bf r}')d^3{\bf r}'$,
where the kernel $Q$ depends on the Fermi surface\cite{nl,book},  the
pairing symmetry\cite{Franz}, and the field orientation.  Here, we
consider cubic or tetragonal s-wave materials in fields along the $c$ axis
so that $Q ({\bf r})$ has the square symmetry.

The kernel $Q(r)$ decays over the nonlocality range  
$\rho = f(T,\ell)\xi_0$, where $\ell$ is the mean-free path for nonmagnetic
scatterers and $\xi_0$ is the BCS zero-$T$ coherence length. The function
$f$ decreases slowly with $T$ and is suppressed strongly by scattering
\cite{nl,book}.  The nonlocality adds to the intervortex interaction a
short-range potential $V(x,y)$ with  the symmetry of the crystal.  In  the
low field limit, $V$ is irrelevant and the VL is triangular; still,
$V$ removes the orientational degeneracy and locks the VL onto certain
crystal direction.  With decreasing intervortex spacing $a(B)$, the 
potential $V $ drives the triangular VL into a square at a field
$H_{\Box}(T)$. The transition  curve $H_{\Box}(T)$ is the subject of
this work.
 
The nonlocal London model describes correctly the observed structure and 
orientation of rhombic VLs in small fields \cite{n8} and the structural
evolution toward the square \cite{vinn}. Moreover, the fast increase of
$H_{\Box}$ with increasing impurity concentration predicted by the model 
has been verified by SANS  \cite{imp}. Still, there is 
an open question on what happens  when the applied field $H > H_{\Box}$ 
keeps increasing. The nonlocal London  model per se implies that the
square VL should persist all the way up to $H_{c2}$ since the shorter the
intervortex distance, the stronger the role of the square-symmetric
potential $V$. This conjecture is supported by calculations based on the 
GL theory extended to include  the 4th order gradient terms
\cite{n5,gl1,gl2}; the model   predicts  that  the lines $ H_{\Box}(T)$
and $H_{c2}(T)$ should cross as sketched in the inset to Fig. 1.

However, recent SANS data for ${\rm LuNi_2B_2C}$ \cite{esk} indicate
that instead of crossing  $H_{c2}(T)$, the line $H_{\Box}(T)$
 curves up to avoid $H_{c2}$ and becomes two-valued, see Fig. 1. 
 The upper branch of $H_{\Box}(T)$ marks the re-entrant  transition of
the rhombic VL in increasing fields, which then evolves toward the
triangular VL as $H\to H_{c2}$. In this Letter we argue that vortex
fluctuations  combined with nonlocal  effects can account for this
unexpected behavior.  
 
Below we evaluate the mean-squared amplitude $\overline {u^2}$ of thermal
vortex fluctuations using the elastic energy of the deformed VL. Unlike
the case of {\it isotropic} superconductors, an important
role in this energy is played by VL {\it rotations}  relative to
the crystal.  We show that $\overline{u^2}$  remains finite at the
transition line $H_{\Box}(T)$, but diverges as $B$ approaches $H_{c2}(T)$.
In the vicinity of $H_{c2}(T)$, the anisotropic nonlocal potential $V(x,y)$
is  averaged out by fluctuations, and the interaction becomes isotropic. 
As a result, the rhombic VL becomes preferable, turning into the 
hexagonal Abrikosov VL as $B\to H_{c2}$. The re-entrance of the rhombic VL
in increasing fields occurs if $\overline{u^2}\sim\xi_0^2 $, i.e.,  the
amplitude of fluctuations needed to wash out the nonlocal effects is much
smaller than that required for the VL melting. This re-entrance,
therefore, can happen in nearly cubic materials, in which vortices are not
split into ``pancakes" and fluctuations are weak.  We evaluate the shape
of the $H_{\Box}(T)$ curve using model parameters   of ${\rm
LuNi_2B_2C}$ for which  in the clean limit $\rho(T)\simeq
(1\div 2)\xi_0$\cite{book,***}.  We reproduce   the shape of this curve
seen in the SANS data \cite{esk}. \\

Let us consider the equilibrium square VL  with 
diagonals  along [100] and [010] directions, which 
we take as $x$ and $y$ axes. We are interested mainly in the high field
region, where the VL can be considered as incompressible \cite{ehb}.
 Then the displacement of vortices $u_i(x,y;z)$ ($i=x,y$)  
satisfy div${\bf u}=\partial_xu_x+\partial_yu_y =0$ and the symmetric
strain tensor 
$u_{ij}=(\partial_ju_i+\partial_iu_j)/2$ has only two independent
components: $u_{xx}$ and $u_{xy}$. Since the VL orientation is locked on
the crystal,   rotations of VL about
[001], $\omega_{xy}=(\partial_xu_y-\partial_yu_x)/2$, cause an energy
increase.   Then the energy density $U$ of the deformed VL reads
\cite{mir}:
	\begin{eqnarray}
	2 U=c_s u_{xx}^2 + c_x u_{xy}^2 +	c_{\omega}\omega_{xy}^2
+c_{44}(\partial_z{\bf u})^2 . 
	\label{f}
	\end{eqnarray}
 Our notation  is motivated by the following. A uniform
displacement $u_x = \epsilon x$,  $u_y = -\epsilon y$ ($ \epsilon\ll 1$ is
a   constant) is of a special interest: it is this deformation
 that transforms the square above $H_{\Box}$ into a rhombus for
$B < H_{\Box}$. This deformation was named ``squash"\cite{kc}, so
that $c_s$ is the squash modulus. Since this is a second order phase
transition,  $c_s$ must vanish  at $B=H_{\Box}$.  Further, the ``simple"
$x$-directed shear is $u_x = \epsilon y$, $u_y=0$ so that only
$u_{xy}\ne 0$; thus the notation $c_x$ for this shear mode.  The shear
energy depends on the shear direction, but the corresponding moduli can
all be expressed in terms of $c_x$ and  $c_\omega$ \cite{mir}.   We take
all moduli in Eq. (\ref{f}) except the tilt $c_{44}$ as practically
nondispersive \cite{ehb}.

Next, we calculate the mean-squared vortex displacement
$\overline{u^2}(T,B)=\overline{u_x^2}+\overline{u_y^2} $.
Writing the total elastic energy in the  Fourier space and utilizing the
equipartition theorem, we obtain \cite{rem1}:
  	\begin{equation}
	\overline{u^2}=\int\frac{qdqdk_zd\phi}{(2\pi)^3}\frac{2T}{q^2
	c(\phi)/4+c_{44}(k_z,q)k_z^2},
	\label{u}
	\end{equation}
where $k_x=q\cos\phi$, $k_y=q\sin\phi$, and
$ c(\phi)=c_{\omega}+   c_x\sin^22\phi+  c_s\cos^22\phi$. With the large
fields tilt modulus $c_{44}(k)=c_{44}(0)/(1+\lambda^2k^2)$, integration
over $k_z$ and $q$ gives $\overline{u^2}=\overline{u_0^2} \eta$. Here
$\overline{u_0^2} \sim T\lambda/a^2\sqrt{c_xc_{44}(0)}$ is  the
mean-squared  displacement for the triangular VL  with $c_{66}=c_x$ and
$c_{44}(0)=B^2/8\pi$ \cite{blat,ehb}. Taking $\overline{u_0^2} $ for a
hexagonal VL in a uniaxial material for ${\bf B}|| c$\cite{ehb}, we arrive
at 
	\begin{eqnarray}
	\overline{u^2}&=&16\sqrt{2}\pi^2\lambda_a\lambda_c\xi_a T\eta
/\phi_0^2s(b)\,,
	\label{uu} \\
	s^2&=&b(1-b)^3\ln(2+1/\sqrt{2b}). 
	\label{s}
	\end{eqnarray}
Here, $\lambda_{a,c}$ are penetration depths,
$b=B/H_{c2}$, and 
 	\begin{equation}
	\eta={2\sqrt{c_x}\over \pi} \int_0^{\pi}\frac{d\phi}{
\sqrt{c(\phi)}}=\frac{4}{\pi}\sqrt{\frac{c_x}{c_x+c_{\omega}}}
	{\bf K}\Bigl(\frac{   c_x-  c_s}{c_{\omega}+   c_x}\Bigr),
	\label{r}
	\end{equation}
where ${\bf K}(m)$ is a complete elliptic integral.

As follows from Eqs. (\ref{uu})-(\ref{r}), $\overline{u^2}(T,B)$ remains
{\it finite} at the instability point $c_s=0$. Thus, vortex fluctuations
do not affect  the mean-field character of the second order transition at
$H_{\Box}$.  This unusual behavior occurs because the rotational
modulus $c_{\omega}$ is {\it finite}, otherwise  $\overline{u^2} $ would 
diverge logarithmically as $B\to H_{\Box}$.

The moduli $c_x$ and $c_{\omega}$ in intermediate fields, $H_{c1}\ll B\ll
H_{c2}$, are  estimated as $c_x\simeq \phi_0 B/(8\pi\lambda)^2$  and
$c_{\omega}\simeq\rho^2B^{2}/2\pi\lambda^2\sqrt{b}$ \cite{app}.  For 
$c_{\omega}/c_x\sim \rho^2/a\xi\ll 1$, we obtain
$\eta\simeq (2/\pi )\ln (16c_x/c_{\omega})$.  Hence, the nonlocality
affects $\overline{u^2}$  via the weakly-varying factor
$\eta\sim\ln(\xi/\rho b^{1/4})$, i.e., $\overline{u^2}(T,B)$ is insensitive
to detailed behavior of $c_{\omega}(T,B)$ and $c_x(T,B)$.
 Since $\overline{u^2}\propto (H_{c2}-B)^{-3/2}$ diverges only at $H_{c2}$,
the fluctuations suppress the weak VL-crystal coupling  mainly  near
$H_{c2}$. \\

 We now turn to evaluation of the curve $H_{\Box}(T)$.   The
free energy $F$ of fluctuating VL can be written as:
 	\begin{equation}
	F=\frac{B^2}{8\pi}\sum_G\frac{e^{-(C\xi^2 + \overline{u^2})G^2/2}}
	{1+\lambda^2G^2+\rho^2\lambda^2G_x^2G_y^2}\,.
	\label{fl}
	\end{equation}
Here the sum  runs over 
$G_x = \pi\sqrt{2}(m-n)/a\tan^{1/2}(\beta/2)$,
$G_y=\pi\sqrt{2}(m+n)\tan^{1/2}(\beta/2)/a$, which form the reciprocal
lattice; 
$\beta$ is the apex angle of the rhombic
cell, and $m,n$ are integers. The last term in the denominator
describes the nonlocal correction to the London theory
\cite{nl,gl1}. The factor $\exp(-C\xi^2G^2/2)$ accounts for the
finite size of the vortex core \cite{ehb}, which provides 
a cutoff in the London model. The  $B$ dependent quantity $C$  was 
estimated  theoretically as $0.5<C<4$  \cite{core}. By and large, the data
on the field dependence of  SANS form-factors are consistent with
$C\simeq1$ \cite{peter}, the value we adopt below.

Fluctuations enter the energy (\ref{fl}) via the Debye-Waller
factor $\exp(-\overline{u^2}G^2/2)$, which accounts for the thermal 
smearing of the vibrating VL\cite{blat}.
For $\overline{u^2}>\rho^2$, the fluctuations wash out the nonlocal 
corrections in $F$, making  the averaged vortex interaction isotropic.
Hence, the re-entrant transition can be driven by thermal vortex
displacements of the order of   $\rho\sim\xi_0$.
 The instability field $H_{\Box}(T)$ thus can  occur well  
below the melting line, which in  materials of interest here is rather
close to $H_{c2}$.

To  find $H_{\Box}(T)$ at which $c_s = \partial^2F/
\partial\beta^2|_{\beta =\pi/2}=0$, we 
differentiate $F$ of Eq. (\ref{fl}) and obtain:
 	\begin{equation}
	\sum_{mn}\frac{e^{-pg}}{d}\Bigl[(2pmn)^2+\Bigl(\frac{8m^2n^2}{d}-g\Bigr)
	\Bigl(p+\frac{1}{d}\Bigr)\Bigr] =0\,.
	\label{cs}
	\end{equation}
Here, $p=[\pi C+ \chi(T)\eta(t, b)/s(b)]b$,
$g=m^2+n^2$, $d=\mu+g+\zeta (b)(m^2-n^2)^2$,
$\mu=1/2\pi b\kappa^2$.
The dimensionless control parameters $\chi$ and $\zeta$ quantify the 
amplitude of thermal displacements and the nonlocal corrections:
	\begin{equation}
	\chi=\frac{16\sqrt{2}\pi^3\lambda_a\lambda_cT}{\phi_0^2\xi_a},
	\qquad
	\zeta =\frac{\pi}{2}b\left(\frac{\rho}{\xi}\right)^2 .
	\end{equation}
Note that nonlocality enters also the exponent $p$ in Eq. (\ref{cs})
via the parameter $\eta$.

For the further analysis, we assume that $\lambda (T)
=\lambda(0)/(1-t^2)^{1/2}$ and $\xi (T)=\xi(0)/(1-t^2)^{1/2}$, 
where $t=T/T_c$ (qualitatively, our results do not change if other 
plausible $T$ dependences are used). 
Then 
	\begin{eqnarray}
	\chi&=&\chi_0 t/\sqrt{1-t^2},	\qquad\qquad\quad 
\zeta=\zeta_0b(1-t^2),
	\label{aob} \\
\chi_0&=&\frac{16\sqrt{2}\pi^3\lambda_a(0)\lambda_b(0)T_c}
{\phi_0^2\xi_a(0)},		\quad
	\zeta_0 =\frac{\pi}{2}\left(\frac{\rho}{\xi_0}\right)^2,
	\label{ab}
	\end{eqnarray}
 In the clean limit $\zeta_0\sim 1$; with increasing   scattering,
$\zeta_0(\ell)$ drops fast \cite{nl}.  For
${\rm LuNi_2B_2C}$ with $T_c=16\,$K, $\xi_0\approx 70$\AA,
$\lambda_a(0)\approx 10^3$\AA, and
$\lambda_c(0)\approx 1.2\times 10^3$\AA, we estimate $\chi_0\approx
6.4\times 10^{-3}\ll 1$. The smallness of $\chi_0$ indicates that   
fluctuations  contribute little to the thermodynamics of stable VLs
 \cite{blat,ehb}. As shown below, being crucial on the upper branch of
$H_{\Box}(T)$,  fluctuations are   negligible on the lower branch. 

 Strictly speaking, $H_{\Box}(T)$ should be calculated
{\it self-consistently} taking into account the effect of fluctuations
on the relevant moduli \cite{scha}.
However, since $\overline{u^2}$  is finite at $H_{\Box}$, we may neglect
the thermal softening of $c_x$  and $c_{\omega}$. Then,
$H_{\Box}(T)$ is just a root of Eqs. (\ref{cs}),
which we  find numerically.
The factor $\eta$ which enters  
$p$ in Eq. (\ref{cs}) is a much weaker function of $t$ and $b$ than 
  $\chi(t)/s(b)$ ($\eta$
varies from 1.6 for $c_{\omega}/c_x=1$ to 3 for $c_{\omega}/c_x=0.1$). 
For this reason we disregard variation of $\eta$, adopting  
  $\eta=2.7$ for ${\rm LuNi_2B_2C}$ with  $c_{\omega}/c_x\simeq
0.2$ at $H_{\Box}$  \cite{mir}.

The results of the numerical solution of Eq. (\ref{cs}) are shown in Fig.
2. It is seen that fluctuations do give rise to the
re-entrant square-to-rhombus transition in high fields, in a 
qualitative agreement with SANS data of Fig. 1.  In fact,
fluctuations  change radically the VL phase diagram in high fields, 
while weakly affecting the low-field branch of $H_{\Box}$.
The difference between $H_{\Box}(T)$ and
$H_{c2}(T)$ is significant (except the low-$T$ clean limit) which
justifies the use of the London model. As the ratio $\rho_0/\xi_0$
decreases (e.g., due to  nonmagnetic impurities),
the region of the square VL on the $H-T$ diagram  shrinks.
The raise of the lower branch of $H_{\Box}(T)$ has been seen on ${\rm
Lu(Co_xNi_{1-x})_2B_2C}$, for which  the mean-free path 
$\ell$ was suppressed by the Co doping \cite{imp}.
 
It is worth noting that although the calculated curves $H_{\Box}(T,\ell)$
reproduce correctly qualitative features of the SANS data of Fig. 1,
actual position of the upper branch is sensitive not only to the accuracy 
with which we know the elastic moduli and the parameters   for their
evaluation, but also to the precise value of the empirical cutoff constant
$C$, see Eq. (\ref{f}). The information  on the actual position
of the upper branch of $H_{\Box}(T,\ell)$ is still scarce, and we hope to
refine our approximations when the data are available.   \\

Now we  comment briefly on the possible effect of the VL transition
on the flux pinning. Since the instability of the square VL at $H_{\Box}$
is not accompanied by divergence of $\overline{u^2} $, the critical current
density $J_c$ evaluated within the collective pinning theory,  
should not be sensitive to the VL transition.  Indeed, the correlation
function of vortex displacements $\langle u({\bf r})u({\bf r'})\rangle$
can be evaluated with the help of Eq. (\ref{u}) in which $T$ is replaced by
$\gamma_p\exp i{\bf k}({\bf r-r'})$, where $\gamma_p$ is the pinning
parameter\cite{blat}. At $H_{\Box}$, the squash modulus vanishes, 
but $\langle u({\bf r})u({\bf r'})\rangle$  remains of the same order as
for a triangular London VL, to the accuracy
of the  weak logarithmic factor $\eta\sim \ln(\xi/\rho b^{1/4})\sim 1$.
Thus, contrary to the claim of Ref. \cite{gl2},   neither the pinning
correlation length nor $J_c$  are significantly affected by the squash
softening near $H_{\Box}$.

 The two-valued $H_{\Box}(T)$ and the re-appearance of the
triangular VL at $H\to H_{c2}$ are  generic features which are not
  limited to nonmagnetic borocarbides.  The
low-$T$ SANS experiments on antiferromagnetic
${\rm TmNi_2B_2C}$ have revealed the  triangular VL near $H_{c2}$, which
  evolves into a square as the field {\it decreases}\cite{es}. At
this stage the effect of antiferromagnetic ordering upon VLs is unclear,
but  fluctuations could certainly contribute to the re-entrant
VL transition in ${\rm TmNi_2B_2C}$ as they do in ${\rm LuNi_2B_2C}$.
Similar behavior was seen in ${\rm YNi_2B_2C}$ \cite{Mona}. Another
candidate for studying effects of vortex fluctuation  is
$\rm V_3Si$, in which the rhombic VL was observed at $T<5^{\circ}$K and  
$H=10\,$kOe ($H_{\Box}\simeq 15\,$kOe); as $T$ increases at the fixed
field, the rhombic VL evolves toward the hexagonal  one as $T\to T_{c2}(H)$
\cite{n7}.\\

In conclusion, we present a model of the structural VL transition at
$H_{\Box}(T)$ affected by thermal fluctuations of vortices.
We show that the curves $H_{\Box}(T)$ and $H_{c2}(T)$ do not cross,
instead $H_{\Box}(T)$ becomes two-valued.  The upper branch of
$H_{\Box}(T)$  corresponds to a re-entrant transition of the
rhombic VL, in accordance with recent SANS observations 
on ${\rm LuNi_2B_2C}$.\\

We are grateful to M. Dodgson and M. Yethiraj for useful discussions.  M.
Eskildsen  kindly provided the data for Fig.1. The work of AG was
supported by the NSF  MRSEC (DMR 9214707) and AFOSR. Ames Laboratory is
operated for US DOE by the Iowa State University under contract No.
W-5405-Eng-82.

\begin{figure}
\epsfxsize= 0.7\hsize  
\centerline{
\vbox{
\epsffile{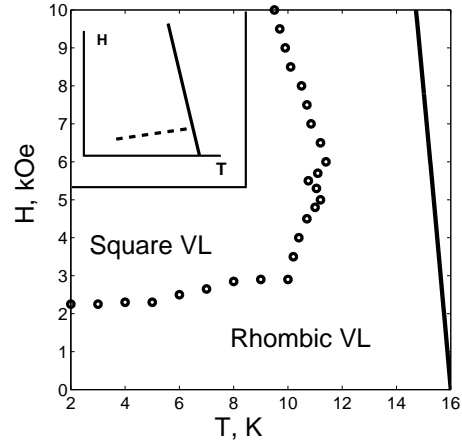} 
}}
\vskip \baselineskip
\caption{The transition line $H_{\Box}(T)$ (circles) in ${\rm LuNi_2B_2C}$ 
observed by SANS [17]. The inset shows    
$H_{\Box}(T)$ (dashed line) predicted by the extended GL theory [6]
without  vortex fluctuations. The solid lines shows $H_{c2}(T)$.
}
\label{fig1}
\end{figure}

\begin{figure}
\epsfxsize= 0.7\hsize  
\centerline{
\vbox{
\epsffile{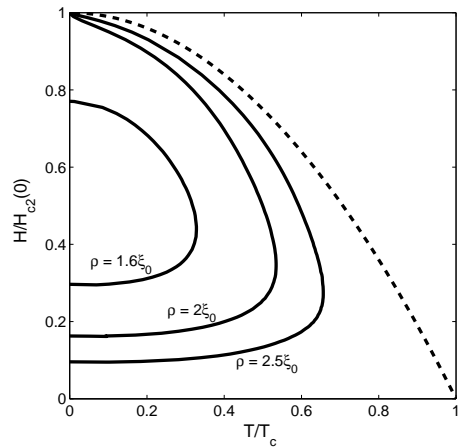}  
}}
\vskip \baselineskip
\caption{The transition lines $H_{\Box}(T)$  obtained by numerically
solving Eq. (\ref{cs}) for  $\chi_0=0.0064$, $C=1$, 
and a few ratios of $\rho/\xi_0$.  The dashed line  is $H_{c2}(T)$. 
}
\label{fig2}
\end{figure}

\end{multicols}

\begin{references}
\bibitem{app}The modulus $c_{\omega}=\partial^2_{\psi\psi}F$ is 
determined by    		
the energy change due to rotation of the VL over the angle
$\psi\to 0$. Substituting $G_x\to G_x\cos\psi - G_y\sin\psi$,
$G_y\to G_x\sin\psi + G_y\cos\psi$ into Eq. (\ref{fl}), we obtain in 
the first order in $\rho^2$:
$c_{\omega}=(B^2\rho^2/4\pi\lambda^2)\sum_{mn}'(1-8m^2n^2/g^2)\exp(-pg)$,
where $g=m^2+n^2$, and $p =\pi C b$.   For $p\ll 1$, the double sum 
is $2\sqrt{\pi /p}$,
thus $c_{\omega}=B^2\rho^2/2\lambda^2\sqrt{p\pi}$.
\end{references}
\end{document}